%% file: main.tex
\documentclass[acmsmall,screen]{acmart}

\AtBeginDocument{%
  }

\usepackage{algorithm}
\usepackage{algorithmicx}
\usepackage{algpseudocode} 
\usepackage{xcolor} 
\usepackage{subcaption}
\usepackage{svg}
\usepackage{booktabs}
\usepackage{blindtext, graphicx}
\usepackage{soul}
\usepackage{xspace}
\usepackage{algpseudocode}  
\usepackage{amsmath} 
\usepackage{indentfirst}
\usepackage{amsfonts}
\usepackage[]{hyperref}
\usepackage{breakcites}
\usepackage[capitalise,nameinlink,noabbrev]{cleveref}
\usepackage{paralist}
\usepackage{listings} 
\usepackage{multirow, makecell}
\usepackage{tcolorbox}
\usepackage{wrapfig}

\usepackage{color}

\tcbset{colframe=black,colback=white,arc=0mm,boxrule=0.4pt,nobeforeafter,tcbox raise base,left=0mm, right=0mm}

\floatname{algorithm}{Algorithm} 

\newcommand{\tool}{\textsc{Re}\textsc{Accept}\xspace}

\definecolor{hlred}{rgb}{0.99, 0.76, 0.8}
\definecolor{comment}{rgb}{0.254, 0.411, 0.882}
\definecolor{ACMPurple}{cmyk}{0.55,1,0,0.15}
\definecolor{ACMRed}{cmyk}{0,0.90,0.86,0}
\definecolor{ACMDarkBlue}{cmyk}{1,0.58,0,0.21}

\hypersetup{colorlinks,
      linkcolor=ACMPurple,
      citecolor=ACMPurple,
      urlcolor=ACMDarkBlue,
      filecolor=ACMDarkBlue}

\begin{document}
\title{\tool: Automated Co-evolution of Production and Test Code Based on Dynamic Validation and Large Language Models}

\author{JianLei Chi}
\email{chijianlei@xidian.edu.cn}
\affiliation{%
  \institution{Xidian University}
  \city{Xi'an}
  \state{Shaanxi}
  \country{China}
}

\author{Xiaotian Wang}
\affiliation{%
	\institution{Harbin Engineering University}
	\city{Harbin}
	\state{Heilongjiang}
	\country{China}
}
\email{wang\_xiaotian@hrbeu.edu.cn}

\author{Yuhan Huang}
\affiliation{%
	\institution{Xidian University}
	\city{Xi'an}
	\state{Shaanxi}
	\country{China}
}
\email{hyh25959@gmail.com}

\author{Lechen~Yu}
\affiliation{%
 \institution{Georgia Institute of Technology}
 \city{Atlanta}
 \state{GA}
 \country{USA}}
\email{lechen.yu@gatech.edu}

\author{Di Cui}
\affiliation{%
	\institution{Xidian University}
	\city{Xi'an}
	\state{Shaanxi}
	\country{China}
}
\email{cuidi@xidian.edu.cn}

\author{Jianguo Sun}
\affiliation{%
	\institution{Xidian University}
	\city{Xi'an}
	\state{Shaanxi}
	\country{China}
}
\email{jgsun@xidian.edu.cn}

\author{Jun Sun}
\affiliation{%
  \institution{Singapore Management University}
  \country{Singapore}}
\email{junsun@smu.edu.sg}

\input{chapters/abstract}




\maketitle
\input{chapters/introduction}
\input{chapters/motivation}

\input{chapters/methodology}

\input{chapters/experiment}
\input{chapters/discussion}

\input{chapters/relatedworks}
\input{chapters/conclusion}
%
%
%
%
%
%
%
\bibliographystyle{ACM-Reference-Format}
\bibliography{references}
\end{document}

%% file: chapters/abstract.tex
\begin{abstract}
Synchronizing production and test code, known as PT co-evolution, is critical for software quality in the software development lifecycle. 
Existing methods for automatic PT co-evolution either utilize predefined heuristic rules or rely on simple application of machine learning techniques.
Due to the limitations of underlying techniques, existing methods either only partially automate PT co-evolution (e.g., only automate obsolete test code identification) or result in low accuracy.

In this paper, we propose \tool{}, a novel approach that leverages large language models and dynamic validation to fully automate PT co-evolution (i.e., capable of both identifying and updating obsolete test cases). 
\tool{} relies on experience-based prompt template generation, dynamic validation, and retrieval-augmented generation techniques to accomplish automated PT co-evolution. 
To evaluate \tool{}'s effectiveness, we extensive experiments with a dataset of 537 Java projects and compared \tool{}'s performance with several state-of-the-art methods. Results show that \tool{} achieved an update accuracy of 60.16\% on correctly identified obsolete test code, surpassing the state-of-the-art technique CEPROT by 90\%.
This confirms that \tool can effectively assist developers in maintaining test code, improving overall software quality and reducing maintenance effort.
\end{abstract}

%% file: chapters/introduction.tex
\section{Introduction}
\label{introduction}
Software testing is an indispensable phase in the software development lifecycle. During testing, developers scrutinize an application's output and performance using a (often sizeable) suite of pre-defined test cases~\cite{pan1999software}.
These test cases may be frequently updated along with the evolution of the application, in order to:
\begin{inparaenum}[a)]
	\item keep existing test cases valid, and
	\item validate the new features in the application.
\end{inparaenum}

Prior studies have demonstrated the importance of synchronizing the test suite with the application~\cite{zaidman2011studying, hurdugaci2012aiding}.
Unfortunately, maintaining such consistency in practice is challenging even for experienced developers. Existing test cases (i.e., \textit{test code}) may not be updated promptly when new changes are committed to the application's code base (i.e., \textit{production code}).
As a result, developers may not be capable of detecting and reproducing bugs incurred by those new changes ~\cite{zaidman2011studying}. 
Researchers have identified multiple factors that hinder an effective co-evolution of production and test code (PT co-evolution for short)~\cite{hurdugaci2012aiding} such as: 
\begin{itemize}
    \item manually maintaining the test suite is time-consuming~\cite{skoglund2004case,runeson2006survey,grindal2006testing};
    \item developers in charge of maintaining the test code may misunderstand the functionality of certain test cases~\cite{engstrom2010qualitative};
    \item after modifying the production code, systematically identifying all affected test cases is challenging due to the complexity of software systems~\cite{buchgeher2013towards}.
\end{itemize}
It is thus important to develop techniques for automatic PT co-evolution, i.e., automatically identifying and updating obsolete test cases and even introducing new test cases whenever the production code is updated. This challenge has recently attracted much research interest in the software engineering areas~\cite{hurdugaci2012aiding, wang2021understanding, sun2023revisiting}.

Existing approaches leverage learning techniques, such as K-Nearest Neighbors (KNN)~\cite{guo2023kape}, Neural Machine Translation (NMT)~\cite{jiang2021cure}, and fine-tuned pre-trained models~\cite{fu2022vulrepair, hu2023identify}, to identify implicit correlations between production code and test code.
Although many approaches aim to identify obsolete test cases, they exhibit significantly low accuracy in updating those cases.
According to previous evaluations on a group of obsolete test cases, CEPROT, a pre-trained model fine-tuned with 4,676 samples, achieved only an accuracy of 35.92\%. Meanwhile, KNN and NMT-based approaches performed even worse, with accuracies of 7.77\% and 22.33\%, respectively~\cite{hu2023identify}.
Such low accuracies may be due to reasons such as insufficient training data, overfitting to specific patterns, or failing to extract general patterns for unknown cases.
Furthermore, learning-based approaches lack mechanisms to interactively and iteratively improve accuracy during the inference phase.
Although reinforcement learning can be applied to guide test repair or obsolete test case updating, it requires expert knowledge to design the model, which may be cumbersome for application developers~\cite{li2019intent}.



Since 2022, Large Language Models (LLMs) have gained significant popularity across multiple research areas. These models, trained on vast amounts of data, outperform traditional deep learning models in code generation tasks~\cite{gu2023llm}.
During code generation, users can interact with an LLM and refine intermediate outputs with additional prompts. This interactive process allows users to incrementally improve the generated code, leading to more accurate and customized results.
In this work, we apply LLMs, as well as, dynamic validation techniques to the PT co-evolution problem, aiming to improve the accuracy of updating obsolete test cases in an incremental and reactive manner.
We propose a novel LLM-based approach, \tool{}, which stands for ``REasoning-Action mechanism and Code dynamic validation assisted Co-Evolution of Production and Test code'' (see~\cref{fig:overview} for \tool{}'s workflow).
\tool{} achieved an average accuracy of 71.84\% on our collected data set, significantly outperforming prior approaches.
All test cases updated by \tool{} have been dynamically validated to confirm correctness in both syntax and semantics. 
In contrast, most prior approaches only compared the updated test cases with the expected results using certain metrics (e.g., BLEU~\cite{papineni2002bleu} and CodeBLEU~\cite{ren2020codebleu}).
Such metrics hardly capture the code semantics, which makes the overall approach ineffective.  

\tool{} is capable of identifying obsolete test cases and then automatically updating them according to the production code.
To improve the accuracy of PT co-evolution, \tool{} leverages ReAct~\cite{yao2022react} mechanism and Retrieval Augmentation Generation (RAG)~\cite{lewis2020retrieval} to interact with the underlying LLM automatically.
\tool{} is designed to minimize the human effort to review intermediate results and steer the LLM.
To gauge the quality of a LLM-generated test case, \tool{} employs a set of third-party tools, including the Java compiler~\cite{javac}, JUnit~\cite{junit}, and JaCoCo~\cite{jacoco}. These tools dynamically validate the test case by checking syntax (Java compiler), verifying semantics (JUnit), and assessing test coverage (JaCoCo).
After the dynamic validation, \tool{} creates new prompts to assist the LLM in refining the test case. This process continues until the test case passes all three validations or a predefined limit is reached.



To evaluate the effectiveness of \tool{}, we constructed a comprehensive dataset (including and extending those from previous work~\cite{wang2021understanding, hu2023identify, sun2023revisiting}), containing 537 projects with 23403 samples. 
We compared the performance of \tool{} with KNN, NMT, and CEPROT.
The evaluation results indicate that overall, \tool{} attained an average accuracy of 71.84\% on the updating tasks, significantly improving existing approaches. Furthermore, \tool{} is tasked to identify and subsequently update obsolete test codes, which achieved an average accuracy of 60.16\%, surpassing KNN's 8.51\%, NMT's 19.83\%, and CEPROT's 31.62\%.
Apart from these comparisons with existing approaches, we also conducted ablation studies, revealing that choosing the right LLMs and adjusting parameters may further improve \tool{}'s performance.
In a nutshell, the evaluation results indicate that \tool{} can effectively automate PT co-evolution and thus reduce developers' effort to maintain the consistency between production code and tests.


In summary, we make the following contributions in this work.
\begin{itemize}
	\item We proposed \tool{}, an automatic approach for PT co-evolution, which combines dynamic validation and LLM prompt engineering. \tool{} effectively accomplishes PT co-evolution with minimized human effort.
        \item We built a dataset comprising 537 projects with 23403 samples to help the community conduct research on PT co-evolution. We have made the replication package publicly available~\cite{ReAccept}.
        \item The evaluation results show that \tool{} achieves an average accuracy of 71.84\% on the updating tasks and an average accuracy of 60.16\% when identifying and updating obsolete test code simultaneously, revealing a significant improvement over prior approaches.
\end{itemize}

The remainder of this paper is organized as follows. 
\cref{sec:motivation} delineates a motivation example and explains the difficulties in PT co-evolution.
\cref{sec:methodology} illustrates the workflow of \tool{}, highlighting the details of both the identification and update phases.
We introduce the conducted evaluations for \tool{} in~\cref{sec:experiment}, and compare its performance with a group of prior work.
We list some related work in~\cref{sec:related_work}, discussing their correspondence with \tool{}.
Finally, in~\cref{sec:conclusion}, we make a summary for this paper with a few future directions.

%% file: chapters/motivation.tex
\section{Motivation Example}\label{sec:motivation}
In the software development lifecycle, production code frequently undergoes numerous version updates, and the corresponding test code is expected to be updated accordingly in a timely fashion. 
This is known as PT co-evolution. 
Failing to achieve PT co-evolution, the obsolete test code may lead to pernicious effects, such as reduced test coverage, higher maintenance costs, and undetected bugs.


\begin{figure}[t]
	\centering
	\includegraphics[width=\columnwidth]{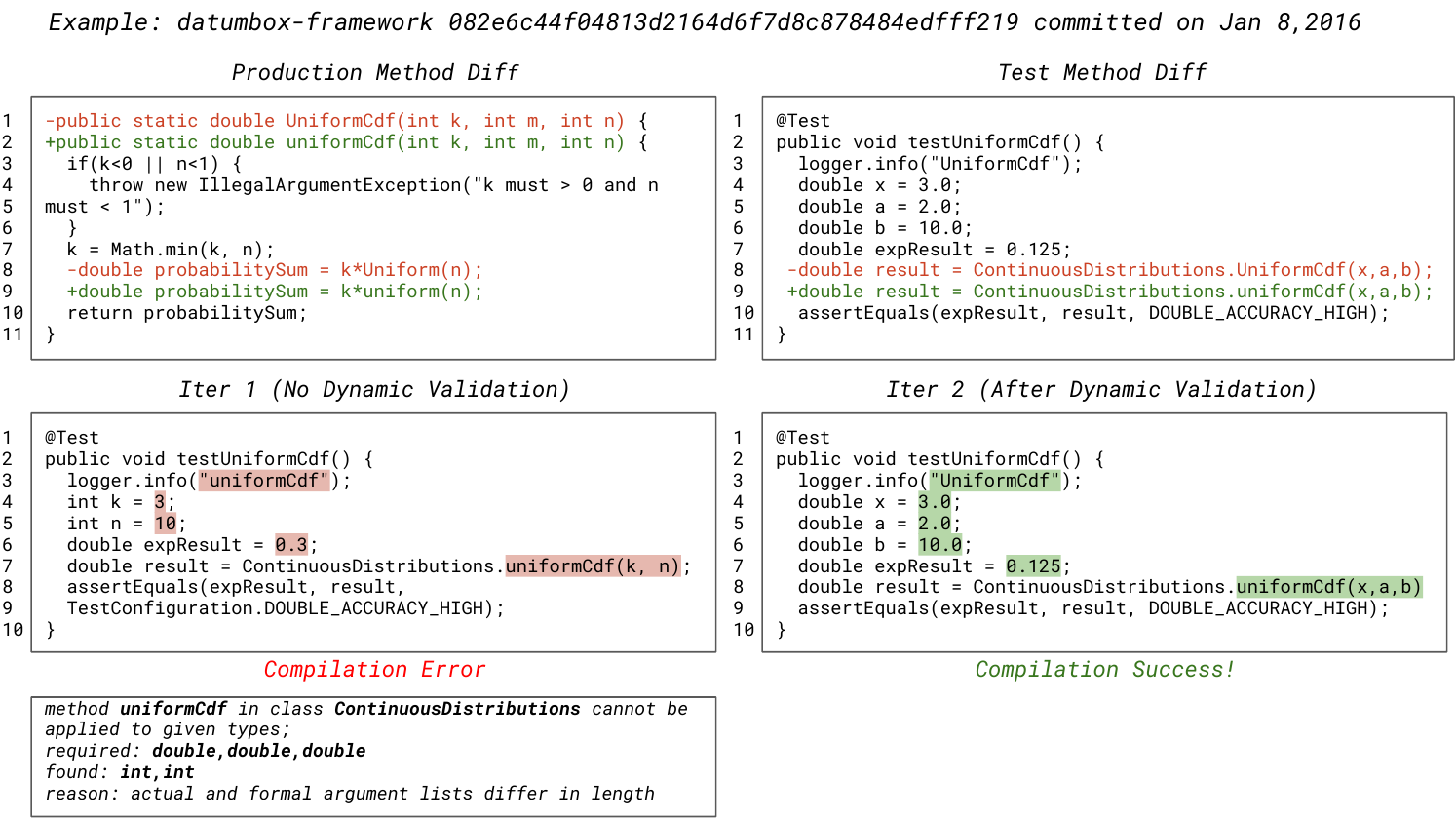}
	\caption{Example of obsolete test code}
    \label{fig:motivation}
\end{figure}

Although maintaining the co-evolution of test code with production code is important, it often presents challenges in practice, particularly in a large software project managed by different teams. 
Previous work~\cite{hurdugaci2012aiding} identified four factors that complicate PT co-evolution in real-world software projects:

\begin{itemize}
	\item Maintaining a test suite is expensive. Given the limited time and resources, developers often deprioritize the co-evolution of test code.
	\item Developers may not be cognizant of all relevant test code related to a specific feature, resulting in the negligence of obsolete test code. 
	\item Even experienced developers may not understand the whole project due to the complexity of modern software systems. When new production code is introduced into the project, the team maintaining the test code may not be able to update it independently.
	\item Manually identifying and updating test code demands extensive domain-specific knowledge, and few tools are available to help developers accomplish the co-evolution of test code.  
\end{itemize}

Aware of such difficulties, we aim to develop a new methodology to facilitate PT co-evolution, ensuring that only verified and correct test code is committed to the code base, thereby greatly reducing the effort required from developers. 
Solely applying LLM to update obsolete test code may be impractical for real-world software projects.
Since LLM may produce undesirable or even erroneous output, developers must continuously validate the generated test code.
To relieve developers from such cumbersome tasks, we propose an approach to help LLM interact with third-party dynamic validation tools automatically, guaranteeing the quality of generated test code.

To illustrate the necessity of dynamic validation when updating test code, we exhibit an example from the datumbox-framework project~\cite{datumbox-framework} in \cref{fig:motivation}.
Developers renamed the method from \texttt{UniformCdf} to \texttt{uniformCdf}, and the invoked utility function \texttt{Uniform} was also refactored to \texttt{uniform}.
Without dynamic validation, the test case updated by an LLM calls \texttt{uniformCdf} with insufficient arguments and incorrect argument type, resulting in a compilation error. Such a test case cannot be added to the repository without further modification. 
By analyzing error messages from dynamic validation tools, such as the Java compiler, the LLM generates a valid test case, resolving inconsistencies in the number and types of arguments.

%% file: chapters/methodology.tex
\section{Our Method}\label{sec:methodology}
In this section, we introduce the workflow of \tool{} and describe the details of its key components.

\subsection{Overall Workflow}
The ultimate of \tool{} is to automate the PT co-evolution completely, i.e., automatically identify and update obsolete test code.
While numerous novel approaches have been proposed in recent years~\cite{wang2021understanding, liu2023drift, hu2023identify, chi2022seqtrans, sohn2022using, shimmi2022leveraging, kitai2022have, sun2023revisiting, lian2024imperfect, liu2024guiding}, most focus primarily on the identification problem, with the task of updating outdated test code remaining cumbersome.
In addition, limited research targeting automatic test code updates suffers from low accuracy, rendering them impractical for real-world software~\cite{hu2023identify}.
 
\begin{figure*}[t]
	\centering
	\includegraphics[width=0.9\textwidth]{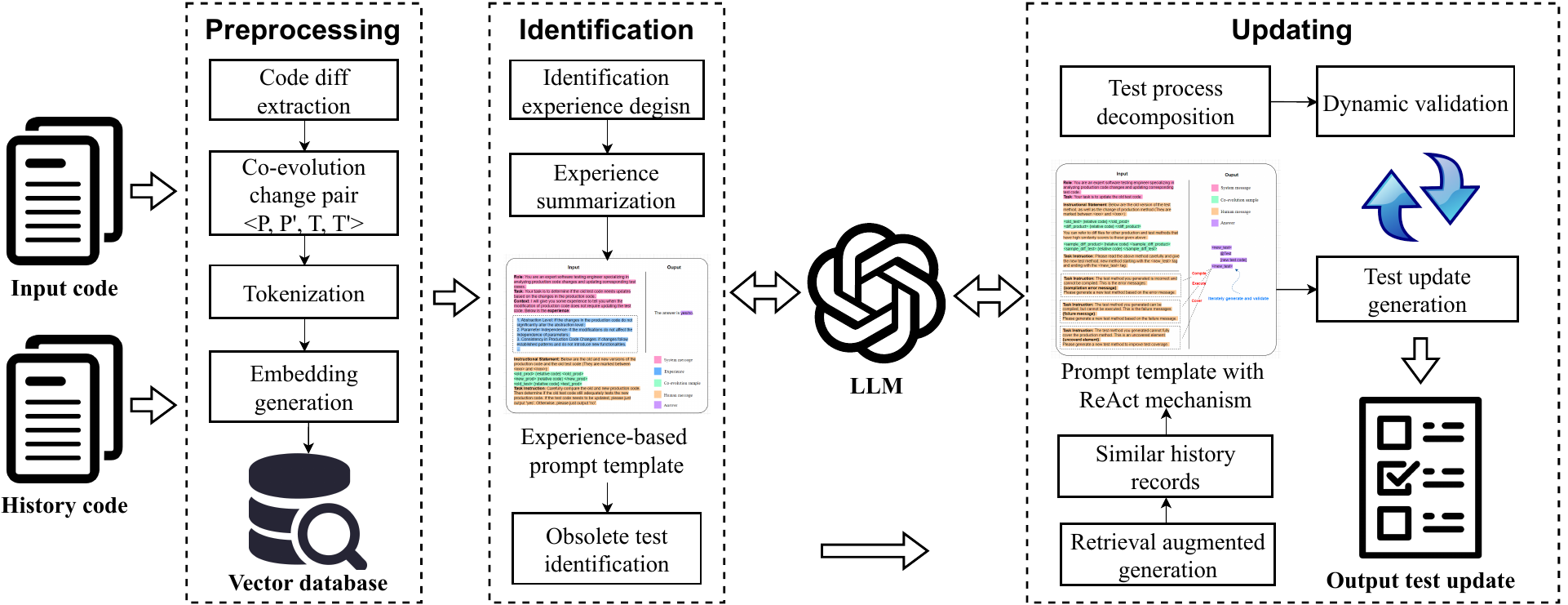}
	\caption{Overview of \tool{}'s workflow}
	\label{fig:overview}
\end{figure*}

Given LLM's impressive capability to process and understand structured text~\cite{ouedraogo2024large, chakraborty2023ranking, yuan2024evaluating, zimmermann2023automating, mezzaro2024empirical, guo2024ft2ra, kou2023model, jin2024can, mu2024clarifygpt, chen2024coder}, \tool{} relies on LLMs and Retrieval Augmentation Generation (RAG), together with dynamic validation, to identify and update obsolete test code precisely.
In \cref{fig:overview}, we illustrate the overall workflow of \tool{}, which consists of three phases: preprocessing, identification, and updating.
The preprocessing phase scrutinizes each project's commit history while extracting per-method code diffs between adjacent commits.
These code diffs are stored in a local vector knowledge base, facilitating prompt construction for subsequent phases.
In the identification phase, \tool{} relies on an LLM with pre-defined experience to determine whether an input code-change pair demands updating.
Finally, in the update phase, \tool{} combines LLM prompting, dynamic validation, and RAG techniques to fix detected obsolete test cases with high accuracy. 

\subsection{Preprocessing}
At this phase, a collection of code commits from real-world projects is organized into structured data, and code change pairs are extracted to construct a local vector knowledge base.

\subsubsection{Data Collection}
We collect samples from 537 Java projects of the Apache Foundation, as well as popular Java projects from GitHub.
All of these projects are managed by Maven, with well-organized repository structures and clear naming conventions. These features make it easier to mine production and test code based on naming conventions.

\tool represents each per-method code diff as a \textit{Change Pair} (CP), comprising four components: $group$ (the project team), $project$ (the project name), $change_p$ (changes in the production code), and $change_t$ (changes in the test code).
\begin{equation}
	CP =(group,project,change_p,change_t)
\end{equation}
$change_p$ and $change_t$ are both 5-tuples of metadata.
\begin{equation}
change_p/change_t = (version,module,package,class,type)    
\end{equation}
These tuples represent the commit hash ID (i.e., SHA1 in Git), the module name, the package name, the class name, and the type of change applied to a single file (either \emph{CREATE}, \emph{DELETE}, or \emph{EDIT}), respectively. 
The values of $group$, $project$, and $version$ together form a unique identifier for the project, while ``$[module]/[package].[class]$'' specifies the relative path of the source file in the project. 


\subsubsection{Knowledge Base Construction}
For each code diff, \tool{} systematically mines the implicit correlation between:
\begin{itemize} 
  \item the original method, $prod_{old}$, and its updated version, $prod_{new}$, 
  \item the original test case, $test_{old}$, and its co-evolved version, $test_{new}$.
\end{itemize}

According to prior studies~\cite{wang2021understanding, sun2023revisiting}, the most common combination of change types is \emph{EDIT-EDIT} (EE), where developers modify both the production and test code. 
Therefore, $change_p$ and $change_t$ are commonly set to \emph{EDIT} in most CPs.

\tool{} employs RAG techniques to retrieve pertinent knowledge from the knowledge base, utilizing a few relevant examples (few-shot) to guide the identification and updating phases. 
RAG enables AI applications to transition from the previous pre-training+fine-tuning mode to the pre-training+prompt mode, greatly simplifying the workload of training models for different demands. 
A crucial task in constructing the knowledge base is representing code changes. 
We aim to reduce redundant information in code changes, shorten prompt words, and improve the performance of LLMs on long sequences. 
If sequence editing is used, it is necessary to explain the meaning of different operations in the editing script to LLMs, which complicates the understanding process. 
Consequently, we differentiate code changes and use code diffs instead of editing scripts~\cite{falleri2014fine} as input. 
To achieve better retrieval results, these diffs are segmented, embedded, and stored in a vector database during the knowledge base construction.

\textbf{Tokenization: }
\tool{} processes word elements in the diff file of the production code and uses LangChain's code segmentation tool to divide it into word element sequences. 
LangChain is a popular framework for building LLM agents by connecting various components~\cite{langchain}. In the code diff, keywords and identifiers are split and converted into tokens, which facilitates representation learning more effectively than statement-level code.
\tool{} employs a block size of 50 with no block overlap.

\textbf{Code Embedding: }
After tokenization, \tool{} vectorizes these word element sequences and uses LangChain's code embedding module to embed such sequences. 
For text embedding, \tool{} applies OpenAI's text embedding model, ada-002~\cite{embedding}. This vector embedding process captures the semantic features of code elements, transforming them into high-dimensional vector representations. These vectors will then be utilized to find knowledge base samples that closely match the target production code changes.

\textbf{Vector Storage: }
The vector data, obtained from code embedding and collaborative modification code pairs, is stored in the vector database Chroma~\cite{chroma} as a dictionary.
That is, it is used as an external expert knowledge base. 
The constructed vector database supports efficient similarity search and retrieval.
It is subsequently used for extracting the most similar sample and guiding the LLMs for code generation.

\textbf{Knowledge Retrieval: }
When a software project receives a change, \tool{} retrieves the most similar sample from the database. 
Then, \tool{} applies this change to the production code and outputs the diff file. The input production code will be transformed into a vector representation. \tool{} relies on \textit{cosine similarity} for sample comparison, which considers the angle between the vectors generated by code changes, and regularizes the code length. The usage of cosine similarity can avoid distance bias due to different code lengths. 
Compared to Euclidean distance, cosine similarity can better capture the semantic information of code changes and thus reflect the semantic similarity of code changes. 
The cosine similarity is calculated using the following formula.
\begin{equation}
	\boldsymbol c\cdot\boldsymbol s=\|\boldsymbol c\|\|\boldsymbol s\|\cos\theta
\end{equation}
\begin{equation}
	\operatorname{similarity}=\cos\theta=\frac{\boldsymbol c\cdot\boldsymbol s}{\|\boldsymbol c\|\|\boldsymbol s\|}=\frac{\sum_{i=1}^{n}c_is_i}{\sqrt{\sum_{i=1}^{n}c_i^2}\sqrt{\sum_{i=1}^{n}s_i^2}}
\end{equation}

Given the production code change $\boldsymbol c$ and the high similarity sample $\boldsymbol s$, the remaining chord $\operatorname {similarity}$ is calculated by the dot product and the vector length.

\subsection{Identification of Obsolete Test Code}
In this phase, the preprocessed code change pairs will be sent to \tool's identifier to determine whether an obsolete test update is needed.
Let $p$ and $p'$ represent the production code before and after the update, with $t$ and $t'$ representing the test code before and after the update. 
The task of obsolete test identification is to define a function $\mathrm{Identify}(p, p', t) $ such that
\begin{equation}
	\operatorname{Identify}(p,p',t)=
	\begin{cases}
		1&\operatorname{if}t\neq t' \text{i.e., $t$ must be updated.}\\
		0&otherwise
	\end{cases}
\end{equation}
Identifying obsolete tests can be regarded as a specialized text classification task, which \tool{} addresses through prompt engineering and RAG techniques.

When the production code undergoes changes and a repair patch is generated, the production code pair $<p,p'>$ and the original test code $t$ are fed into the LLM. 
The model then attempts to understand and capture the semantic relationships in the code modifications, thereby determining the necessity of updating $t$. 
If the update is indispensable, \tool{} compares and searches for the most similar test code update record in the vector knowledge base, to guide the subsequent obsolete test update.

The overall process of obsolete test identification is made up of three steps, including \textit{identification experience learning}, \textit{prompt template design}, and \textit{identification result generation}.

\subsubsection{Identification Experience Learning}
In a preliminary experiment, we directly asked the LLM to identify the obsolete test code only based on $prod_{new}$, $prod_{old}$ and $test_{old}$.
The performance proves to be unsatisfactory.
We observed that the LLM exhibited a cognitive bias when analyzing the test code, presuming that any modification to the production code would render the correlated test cases obsolete.
There might be two probable reasons.
\begin{itemize}
	\item \textbf{Data and corpus deviation}. LLMs are typically trained on vast amounts of text data, which often includes extensive discussions about code updates and maintenance. In many real-world development scenarios, modifications to the production code frequently accompany updates to the test code. This pattern may dominate a significant portion of the training data, causing the model to develop such a bias.
	\item \textbf{Lack of domain-specific knowledge}. Despite possessing a broad spectrum of general knowledge, LLMs may lack expertise and experience in specific domains. There is a tendency to rely on general knowledge, potentially overlooking the nuance between real-world projects.
\end{itemize}

To address these issues, we utilize the LLM to generate a set of generalized experience derived from $CPs$. 
These learned experience provide contextual clues during the identification phase, enhancing identification accuracy. 
In generating these experience, we select samples from the training set that reflect key patterns in production code modifications and their corresponding test code updates. 
This selection preserves a representative diversity, enabling the LLM to learn generalized experience across a wide range of scenarios.


\subsubsection{Prompt Template Design}
\begin{wrapfigure}{r}{0.6\columnwidth}
	\centering	 
        \includegraphics[width=0.6\columnwidth]{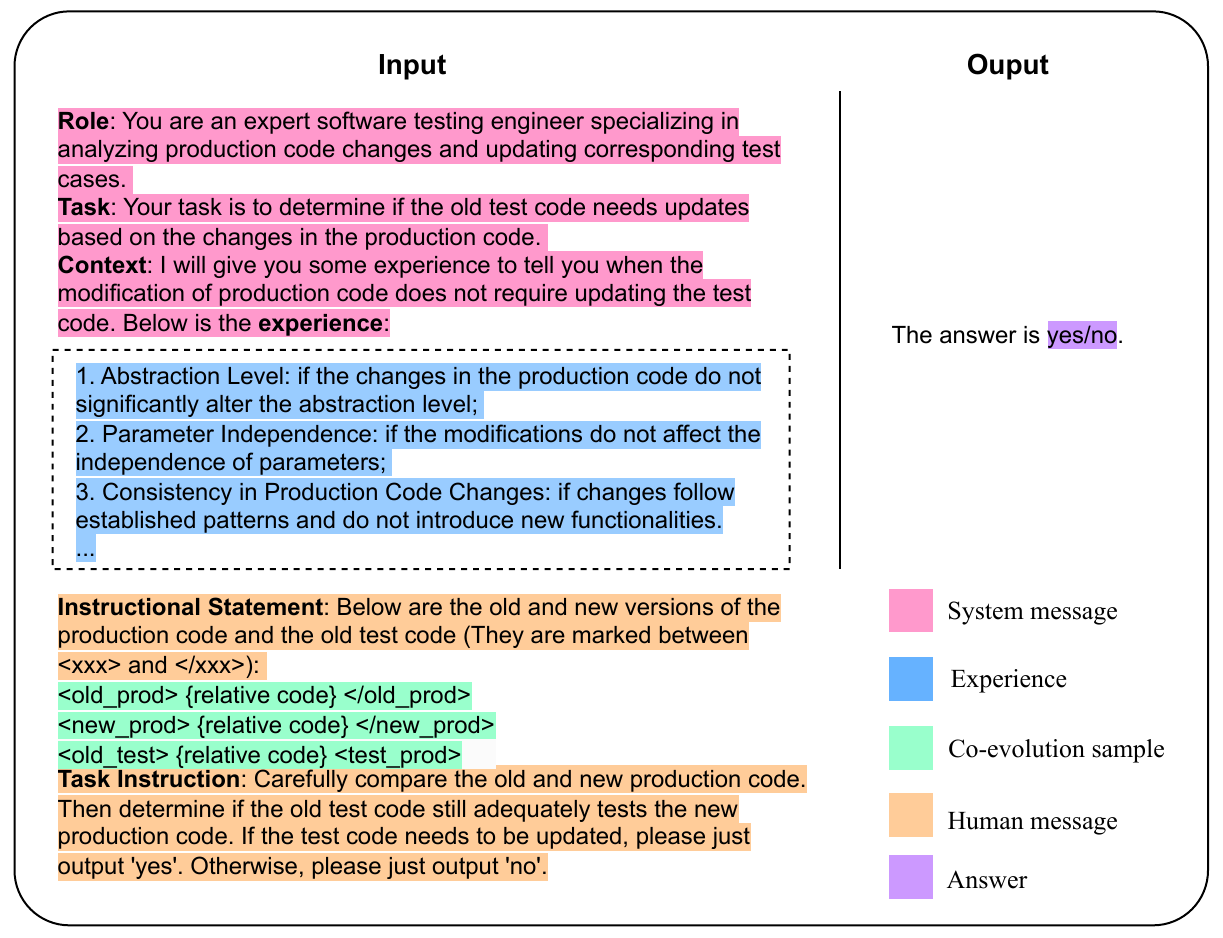}
	\caption{The prompt template of the identification task}
	\label{fig:identification_prompt}
\end{wrapfigure}

Prompt engineering plays a crucial role in identifying obsolete test code. 
By carefully designing the prompts, the LLM can make more accurate judgments. \cref{fig:identification_prompt} illustrates an example of a prompt template we used for the obsolete test identification task. 
The prompt template is primarily composed of two parts: $system\ message$ and $human\ message$. The system message defines the role the LLM should assume and its general behavior. The human message, also known as the user message, includes prompts from the user and may provide examples, historical prompts, or specific instructions for the assistant.
The generalized experience we summarized is also included in the human message. Next, we will provide a detailed explanation of how this experience is generated.

In order to help the LLM accurately determine when the test code needs updating, \tool{} encapsulated a set of experience into a structured format. 
Each experience represents an abstract principle, such as Abstraction Level ($AL$),  Parameter Independence ($PI$), and Consistency in Production code changes ($CP$).

For $AL$, ReAccept identifies cases where significant abstraction changes in the production code indicate a need for corresponding test updates. 
For instance, if a procedural code is refactored into an object-oriented structure, this experience prompts the model to consider updating the related tests to reflect the higher abstraction.
In $PI$, the experience includes scenarios where parameter modifications—such as reordering or replacement—do not impact their independence, meaning tests do not require updates. 
This pattern allows us to avoid unnecessary test changes, maintaining stability when parameter dependencies remain intact.
$CP$ implies that, as long as modifications adhere to pre-existing patterns without introducing new functionality, the test code remains valid. An example would be minor refactoring within an established function that does not alter its core behavior.
Consequently, we obtained a set of ($AL_i$, $PI_i$, $CP_i$) as the experience.

Then, \tool summarizes and updates the experience in the prompt.
Through multiple rounds of iterative optimization, the prompt continuously takes more scenarios into account.
Additionally, we implemented a feedback mechanism in the design of \tool{}. 
We analyzed the underlying causes of erroneous identification results and adjusted the prompt content based on the identification results and the actual performance of the LLM.

\subsubsection{Identification Result Generation}
After multiple rounds of optimization and experience accumulation, our prompt design can now accurately guide the LLM in identifying the obsolete test code. 
During the process of generating identification results, the LLM analyzes the given production and test code based on the experience and guidance in the keywords, and provides a judgment on whether the test code needs to be updated. 
Additionally, the LLM will offer detailed explanations to help developers understand the reasoning behind its judgments.

By continuously optimizing the design of prompts, combined with the powerful natural language processing capabilities and domain-specific experience of LLM, we have significantly improved the accuracy of identifying obsolete test code, laying a solid foundation for subsequent test code updates.

\subsection{Obsolete Test Code Update}
In this phase, \tool updates the detected obsolete test code.
The task of obsolete test update can be described as a function $\mathrm{Update}(p, p', t) = t'$ which returns a test $t'$ that is compilable, testable, and covering updated code in $p$.
In our approach, \tool solves this task in two steps: \textit{test update generation} and \textit{test update validation}. 
Although \tool{} focuses on Java applications, the techniques behind \tool{} are applicable to other programming languages. 

\subsubsection{The Decomposition of Obsolete Test Update}

\begin{figure}[t]
	\centering
	\includegraphics[scale=0.6]{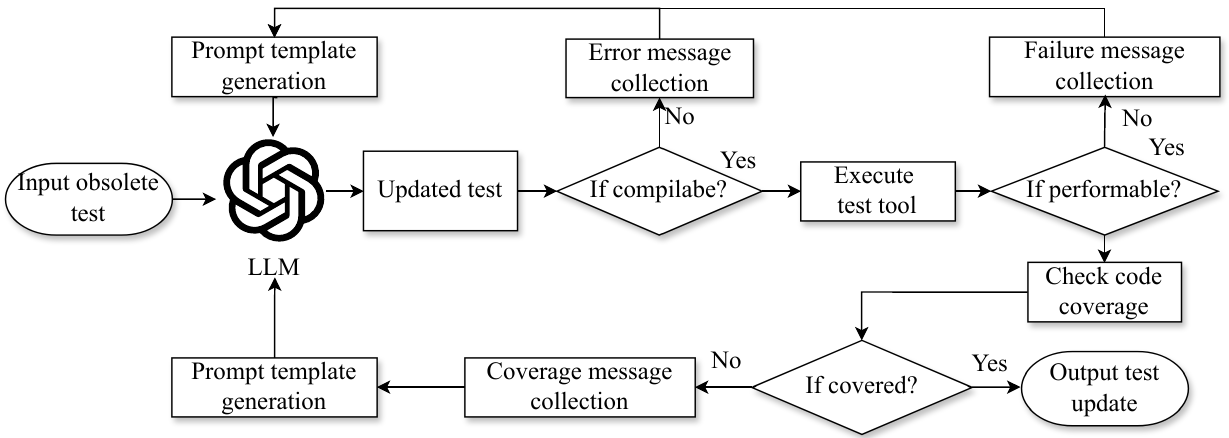}
	\caption{Test process decomposition}
	\label{fig:decomposition}
\end{figure}

As illustrated in \cref{fig:decomposition}, the identified obsolete test code will be sent to \tool's updater. 
The updater combines the obsolete test code with the retrieved most similar historical records to form a few-shot prompt, thereby generating the new test code through the LLM. 
Using dynamic validation, the updater adjusts the output from the LLM to make sure the updated test code satisfies some pre-defined requirements.
First, the test code must be correctly compiled by the Javac compiler~\cite{javac}.
In addition, \tool{} reruns the test code and acquires their coverage to the production code through JaCoCo~\cite{jacoco}.
As a valid update, the generated test code must cover the modified statements in the production code while not arising any runtime errors~\cite{ivankovic2024productive}.

The decomposition of the update phase facilitates the construction of prompt templates for different scenarios, assisting the LLM in fixing erroneous snippets in the test code. 
Due to the inherent variability of testing effects, \tool measures testing code quality using four levels: \textit{compilation failure}, \textit{test failure}, \textit{coverage failure}, and \textit{satisfying all testing requirements}.
This also requires \tool to parse the results of the dynamic validation and interact with the LLM for future test code generation.

\subsubsection{Test Update Generation}

The prompt templates play a pivotal role in guiding update tasks. 
It is also necessary to optimize the prompt template to ensure that the LLM outputs well-formatted text~\cite{peng2024domain}.

\begin{wrapfigure}{r}{0.6\columnwidth}
	\centering	 
        \includegraphics[width=0.6\columnwidth]{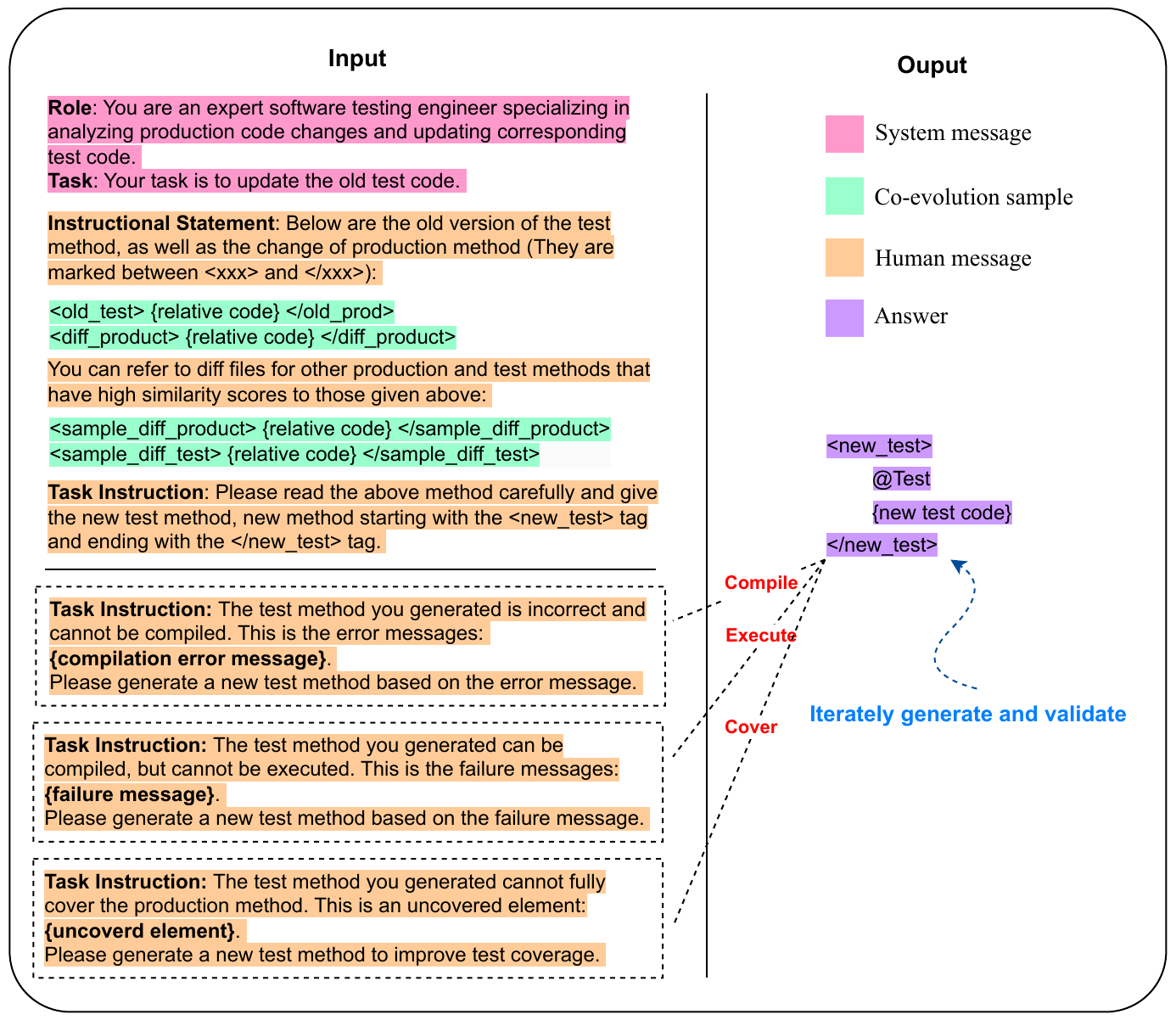}
	\caption{The prompt template of the updating task}
	\label{fig:update_prompt}
\end{wrapfigure}

As has been shown in \cref{fig:update_prompt}, the input parameters accepted by the prompt template for updating tasks are production code before and after the change ($prod_{old}$, $prod_{new}$), the original test code $test_{old}$, and retrieved samples with highest similarity ($prod_{samlpe}$, $test_{sample}$). 
We choose to embed a sample's code diff into the prompt templates rather than directly entering all related production and test code to the LLM.
Since the diff file concentrates on the code changes, it is instrumental in compressing the prompt length and avoiding potential performance downgrade when the LLM encounters a lengthy prompt.

If the test code does not meet the testing requirements, \tool{} uses the test results obtained from feedback information.
\cref{fig:update_prompt} illustrates the three types of feedback templates we constructed. These templates will be integrated into the human message to build a new prompt, which will then be input into the LLM.

\subsubsection{Applying ReAct Mechanism}
After generating prompt templates, \tool initializes the LLM with a conversation memory, to build an agent following the ReAct mechanism~\cite{yao2022react}. 

The agent generates inference paths and specific task behaviors in an interleaved manner. The inference paths facilitate the deduction of behavior plans and the handling of corner cases. 
Meanwhile, the task behaviors allow the agent to access external sources, such as the knowledge base to obtain knowledge related to the current task, i.e., high-similarity samples.
The ReAct mechanism can coordinate LLM to obtain testing information from the actual project environment, which is used to guide LLM to correct compilation errors, testing failures, or low coverage of testing code.


\subsubsection{Test Update Validation}
After constructing the agent, the change patterns ($CPs$) will be input into the agent to predict the updated test code. 
We will also retrieve the most similar sample from the knowledge base to construct a prompt. 
This prompt will then be input into the LLM, allowing us to extract the updated test code from the LLM's output. 
The updated test code will be used to test the modified production code in the environment, generating test results and messages.

\tool{} will revise and dynamically validate the output code through the following process. 
All error messages encountered during dynamic validation will be analyzed to inform future test code generation.

\begin{enumerate}
	\item Update the test code in the original project environment.
	\item Compile the test code to check if the test class file is compilable. 
	\item Rerun the test code, examining whether the execution raises any failures or runtime errors.
	\item Apply JaCoCo to check the coverage at the statement level. The updated test code should cover the changes revealed in diffs of the production code.
\end{enumerate}

The dynamic validation process will keep executing the process shown in \cref{fig:decomposition} until generating a valid test code or reaching a pre-defined cutoff (currently, we empirically set the maximum rounds of iteration to 8).
When the test code passes the checks of compilation, re-execution, and test coverage, \tool{} 
will stop further interaction with the LLM and output the updated test code.

%% file: chapters/experiment.tex
\section{Experiment}\label{sec:experiment}
Our method \tool has been implemented based on the LangChain framework, using OpenAI's gpt-4-0125-preview LLM. The default parameters are set as follows: the temperature of the LLM is set to 0; the top P is set to 1; the frequency penalty is set to 0; the presence penalty is set to 0; and the short-term memory of the conversation chain adopts the context window strategy with a window size of 3.

In this section, we design multiple experiments to evaluate the effectiveness of \tool{}. 
We start by outlining the research questions that guided the evaluation design, followed by a description of the experimental setup and an in-depth discussion of the results.

\subsection{Research Questions}
We aim to answer the following research questions:
\begin{itemize}
    \item \textbf{RQ1:} How effective is \tool on solving the PT co-evolution problem, i.e., identify and update obsolete test code?
    \item \textbf{RQ2:} How effective is \tool on identifying obsolete tests?
    \item \textbf{RQ3:} How effective is \tool on updating obsolete tests?	
    \item \textbf{RQ4:} What are the factors that can impact the performance of \tool?
\end{itemize}

RQ1 aims to provide an overall evaluation of our method, treating the identification and update phases as two parts of a holistic process.
RQ2 and RQ3 then analyze the performance of our method in each phase separately.
At last, RQ4 conducts multiple ablation studies to evaluate the impact of different design choices and parameters on \tool{}'s performance.

\subsection{Experiment Settings}

\subsubsection{Dataset Construction}
With non-trivial effort, we build a comprehensive PT co-evolution benchmark dataset, expanding the datasets from previous work~\cite{wang2021understanding, hu2023identify, sun2023revisiting}, to facilitate our evaluations and future research on PT co-evolution. The dataset primarily consists of Java projects from the Apache Foundation, along with several highly-starred Java projects on GitHub. 
All of these projects are managed using Maven and follow a good structure and naming convention, making it easier to mine paired production and test code based on these conventions. 
In total, the dataset includes 537 projects. The detailed information of the dataset is shown in \cref{tab:dataset}.

We split the dataset into a 90\% training set for the LLMs to gain experience and a 10\% test set to assess the effectiveness of our PT co-evolution approach. 
The co-evolution samples in the dataset were labeled as either positive or negative. 
Positive samples indicate that the corresponding test code requires updating, while negative samples do not require any changes. 
Consequently, only the positive samples from the training set were included in the knowledge base, which can be retrieved using the RAG technique.


Note that \tool conducts dynamic validation in the test code updating phase, which requires setting up the correct test environment (e.g., JDK, third-party packages) for all samples. Thus, to evaluate the effectiveness of updating obsolete test code, we scrutinized six popular Java projects from our dataset that are managed by Maven, examining the test environment of each collected commit.
Details of these projects are presented in~\cref{tab:RQ3testdata}. The first two columns record each project's name and the number of collected commits, respectively. The third column is the number of collected samples, indicating that a single commit may contain multiple samples. 
The number of commits and samples are expressed as ratios, with the numerator representing the count of successfully built and runnable instances.
Overall, our evaluations utilized 74 commits, which included 103 co-evolution samples.
\begin{table}[t]
\begin{minipage}[b]{0.5\columnwidth}
  \centering
  \caption{Data Set Details}
  \label{tab:dataset}
  \begingroup
  \setlength{\tabcolsep}{10pt}
  \renewcommand{\arraystretch}{0.9} %
  \begin{tabular}{cccc}
    \toprule
	\multirow[b]{1}{*}{Data} & \multirow{2}{*}{Pos.} & \multirow{2}{*}{Neg.} & \multirow{2}{*}{Total}\\
    \multirow[t]{1}{*}{Set} & & & \\
	\midrule
	Train & 4676 & 16496 & 21172\\
	Test & 520 & 1771 & 2231\\
	\midrule
	Total & 5196 & 18267 & 23403\\
	\bottomrule
  \end{tabular}
  \endgroup
\end{minipage}%
\hfill
\begin{minipage}[b]{0.5\columnwidth}
  \centering
  \caption{Benchmark Details for Dynamic Evaluation}
  \label{tab:RQ3testdata}
  \begingroup
  \setlength{\tabcolsep}{10pt}
  \renewcommand{\arraystretch}{0.9} %
  \begin{tabular}{lcc}
	\toprule
	Project& Commits & Samples\\
	\midrule
	springside4&2/2&3/3\\
	commons-lang&5/5&8/8\\
	dddlib&2/4&8/10\\
	datumbox&7/7&9/9\\
	openmrs-core&28/34&29/37\\
	basex&30/42&46/63\\
    \midrule
	Total&74/94&103/130\\
	\bottomrule
  \end{tabular}
  \endgroup
\end{minipage} 
\end{table}

\subsubsection{Baselines}


For baseline comparisons, we compared \tool with general methods such as Random Guess (RG), K-Nearest Neighbor (KNN), Neural Machine Translation (NMT), as well as state-of-the-art approaches including SITAR~\cite{wang2021understanding}, CHOSEN~\cite{sun2023revisiting}, and CEPROT~\cite{hu2023identify}.
Since RG, SITAR, and CHOSEN are only capable of identifying the obsolete test code (RQ2), we omitted them when comparing the overall performance (RQ1) and the effectiveness of updating the obsolete test code (RQ3).

\subsubsection{Evaluation Metrics}
When evaluating obsolete test code identification, we measured each approach's accuracy, precision, recall, and F1 score on our dataset. We selected these metrics because the task can be modeled as a binary classification problem, and these metrics are commonly used to assess the performance of classification methods.

For the task of obsolete test code update, we need to compare the test code generated by all approaches with that of the ground truth. We adopted CodeBLEU~\cite{ren2020codebleu} as the evaluation metric. 
As an extension to BLEU~\cite{papineni2002bleu}, CodeBLEU is a commonly used metric for evaluating code generation tasks. 
Prior work~\cite{ren2020codebleu} has revealed that BLEU does not take into account the correctness of programming language syntax and logic, likely to result in syntax or logic errors with high n-gram accuracy. 
Therefore, CodeBLEU is more suitable for our evaluations.


However, CodeBLEU still cannot fully reflect the effectiveness of the updated test code in practice. In~\cite{hu2023identify}, Hu \emph{et al.} show that even test code with a high CodeBLEU score may fail to execute correctly in real projects. The compilation success rate is only 48\%, and the actual correctness rate of the update is only 12.3\%. 
To address the problem, we further complemented CodeBLEU with the following three metrics measured based on dynamic evaluation, when comparing \tool with those baseline approaches. 
\begin{itemize}
	\item Compile Success Rate (CSR): the percentage of test code that is successfully compiled (after building the corresponding project version).
	\item Test Pass Rate (TPS): the percentage of updated test code that successfully runs and passes.
	\item Update Coverage Rate (UCR): the percentage of updated test code that successfully runs, passes, and covers the changes in the product code. 
\end{itemize}
Note that for all rates, higher values indicate better performance.

At last, to evaluate the overall effectiveness of each technique for PT co-evolution, we measured the percentage of correctly identified test code that successfully compile, run, and cover the updated production code.

\subsection{Experimental Results}
\subsubsection{RQ1: The effectiveness on solving the PT co-evolution problem}
\begin{wrapfigure}{r}{0.5\columnwidth}
	\centering
	\includegraphics[width=0.5\columnwidth]{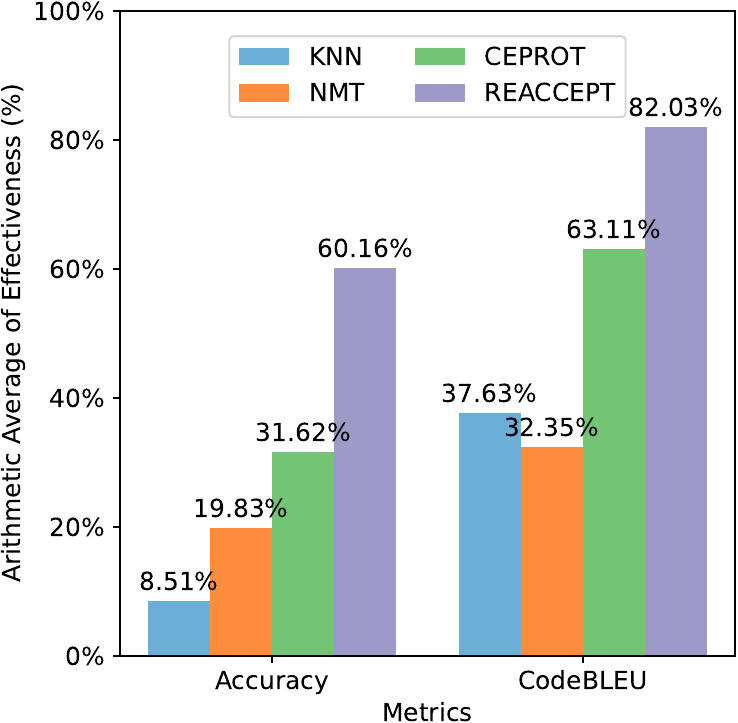}
	\caption{The overall effectiveness of different approaches}
    \label{tab:two_phase}
\end{wrapfigure}

\cref{tab:two_phase} shows the results when \tool{} and baseline approaches aim to solve the PT co-evolution problem completely, i.e., identifying and updating the tests automatically. 
We can observe that \tool outperforms all baseline approaches, showing significant improvements across all metrics.
\tool achieves an update accuracy of 60.16\% on correctly identifying the obsolete test code, surpassing KNN's 8.51\%, NMT's 19.83\%, and CEPROT's 31.62\% by 607\%, 203\%, and 90\%, respectively.
Regarding the CodeBLUE metric, \tool also outperforms other approaches with a score of 82.03\%. 
This yields improvements of 118.7\%, 154.4\%, and 30.4\% over KNN's 37.63\%, NMT's 32.35\%, and CEPROT's 63.11\%, respectively.



Specifically, in the identification phase, \tool identifies more obsolete test code compared to KNN, NMT, and CHOSEN classifiers.
Although CEPROT surpasses \tool in the obsolete test identification phase, \tool demonstrates significantly better performance in the obsolete test update phase, ultimately securing a notable overall advantage.
Note that both phases of \tool are flexible and independent, allowing for the integration of more effective obsolete test identification techniques to further enhance \tool's performance.
Overall, the results suggest that \tool is a useful tool that can significantly reduce the required effort for the PT co-evolution problem.

\subsubsection{RQ2: The effectiveness of the obsolete test identification} 
The above results suggest that \tool is the best approach overall. 
Next, we evaluated the performance of \tool in the two phases separately and compared it to the corresponding state-of-the-art approaches.
\cref{tab:identification_samples} provides detailed statistics on \tool{}'s performance during the obsolete test identification phase, including True Positives, False Positives, True Negatives, and False Negatives.
Next, \cref{tab:identification_performance} highlights the performance difference between \tool and other baseline approaches (RG, KNN, SITAR, NMT, CHOSEN, and CEPROT) using those predefined metrics, such as precision, recall, and F1-score, for both positive and negative samples. Note that accuracy reflects the overall correctness of an approach’s predictions, irrespective of class labels, and is therefore reported as a single value in \cref{tab:identification_performance}.

It can be observed that \tool significantly outperforms RG, KNN, NMT, and SITAR, marginally outperforms CHOSEN, and shows comparable performance to CEPROT. Notably, \tool excels in the negative class, with a precision of 98.92\%, recall of 98.63\%, and F1 score of 98.77\%. 
We found that misidentification mainly results from a lack of comprehensive experience covering all complex code change scenarios. Improving the quantity and quality of the training set can help mitigate such misidentification.
Additionally, utilizing more advanced LLM models could further enhance \tool's identification performance.
We would like to point out that, unlike \tool, which is designed to fully address the PT co-evolution problem, some approaches (such as SITAR and CHOSEN) only partially address the issue and do not automatically update the test code. 
Extending these approaches to incorporate update functionality is also cumbersome due to their complex software infrastructure.
Furthermore, while CEPROT performs well in identifying the obsolete test code, it turns out to be ineffective in updating those test cases, as we will show in the following research question. Overall, the results show that \tool can effectively identify the obsolete test code.

\begin{table}
  \begin{minipage}[b]{0.35\columnwidth}
    \centering
    \caption{\tool's identification result of 2231 Samples}
    \label{tab:identification_samples}
    \begingroup
    \setlength{\tabcolsep}{3pt}
    \renewcommand{\arraystretch}{0.9} %
    \begin{tabular}{lccc}
        \toprule
        \multirowcell{2}[0pt][l]{Identification \\ Labe} &  \multicolumn{2}{c}{Prediction} & \multirowcell{2}{Total}\\
        \cmidrule(lr){2-3}
         & Pos. & Neg. & \\
        \midrule
        Positive & 496 & 24 & 520 \\
        Negative & 53 & 1658 & 1711 \\
        \midrule
        Total & 549 & 1682 & 2231 \\
        \bottomrule
    \end{tabular}
    \endgroup
  \end{minipage}%
  \hfill
  \begin{minipage}[b]{0.62\columnwidth}
    \small
    \centering
    \caption{Effectiveness of different techniques on the obsolete test identification task}
    \label{tab:identification_performance}
    \begingroup
    \setlength{\tabcolsep}{2pt} 
    \renewcommand{\arraystretch}{1.2} 
    \begin{tabular}{l@{\hspace{2pt}}ccccccc}
    \toprule
    \multirow{2}{*}{Method} & \multirow{2}{*}{Acc.} & \multicolumn{3}{c}{Positive}  & \multicolumn{3}{c}{Negative} \\
    \cmidrule(lr){3-5} \cmidrule(lr){6-8}
    & & Prec. & Rec. & F1. & Prec. & Rec. & F1. \\
    \midrule
    RG & 48.68\% & 46.39\% & 49.72\% & 48.00\% & 51.08\% & 47.74\% & 49.35\%  \\ 
    KNN & 74.73\% & 83.40\% & 72.30\% & 77.50\% & 90.56\% & 75.41\% & 82.29\%  \\ 
    SITAR & 84.17\% & 78.30\% & 38.90\% & 52.00\% & 84.83\% & 96.89\% & 90.38\%  \\ 
    NMT & 91.50\% & 82.00\% & 78.85\% & 80.39\% & 94.05\% & 95.09\% & 94.51\%  \\ 
    CHOSEN & 92.89\% & 89.29\% & \textcolor{red}{96.69\%} & 92.84\% & 96.74\% & 89.45\% & 92.95\% \\ 
    CEPROT & 97.50\% & \textcolor{red}{98.30\%} & 90.00\% & 94.00\% & 97.20\% & \textcolor{red}{99.60\%} & 98.40\% \\ 
    \textbf{\tool{}} & \textcolor{red}{98.13\%} & \textcolor{blue}{95.57\%} & \textcolor{blue}{96.49\%} & \textcolor{red}{96.01\%} & \textcolor{red}{98.92\%} & \textcolor{blue}{98.63\%} & \textcolor{red}{98.77\%}  \\ \bottomrule
    \end{tabular}
    \endgroup
  \end{minipage} 
\end{table}

\subsubsection{RQ3: The effectiveness of the obsolete test update.}
Apart from the identification phase, we also evaluated \tool's performance in terms of updating the identified obsolete test code. 
The corresponding evaluation results are shown in~\cref{tab:dynamic_performance}.
We can observe that \tool significantly outperforms all the baselines, in terms of all three metrics.
The CSR of \tool is 85.44\%, which is substantially higher than the best baseline's 49.52\%.
Furthermore, the updated test code generated by \tool have a 71.84\% chance of covering the updated product code, considerably higher than the best baseline's 35.92\%.
These results indicate that most of the test code generated by \tool are not only syntactically correct but also appropriately validate the production code changes, while existing techniques struggle to effectively update the test code.
Overall, \tool achieved an improvement of 825\%, 222\%, and 100\% over KNN, NMT, and CEPROT, respectively.

It is apparent that prior work has largely overlooked the challenge of updating the obsolete test code, possibly due to its complexity.
KNN and NMT are general-purpose methods that can be applied to various tasks, but their performance in PT co-evolution is not satisfactory enough.
CEPROT exhibits reduced accuracy when tackling test code containing numerous statements, as the lengthy test code must be split into multiple smaller code blocks and updated separately.
The underlying issue, as noted by the authors, is that their Transformer-based model is limited in handling long sequence tasks~\cite{hu2023identify}.
On the other hand, our approach \tool demonstrates significant improvements over the state-of-the-art approaches when tackling such lengthy test code. 
\begin{table}
  \small
  \centering
  \caption{Dynamic performance of different techniques on the obsolete test update task}
  \setlength{\tabcolsep}{1.3pt}
  \begin{tabular}{lcccccccccccc}
      \toprule
      \multicolumn{1}{l}{\multirow{2}{*}{Approach}} & \multicolumn{3}{c}{KNN} & \multicolumn{3}{c}{NMT} & \multicolumn{3}{c}{CEPROT} & \multicolumn{3}{c}{\tool{}} \\
      \cmidrule(lr){2-4} \cmidrule(lr){5-7} \cmidrule(lr){8-10} \cmidrule(lr){11-13}
      & CSR & TPS & UCR & CSR & TPS & UCR & CSR & TPS & UCR & CSR & TPS & UCR\\
      \midrule
      springside4 & 66.67\% & 66.67\% & \phantom{0}0.00\% & 33.33\% & 33.33\% & 33.33\% & 66.67\% & 66.67\% & 66.67\% & 100.00\% & 100.00\% & 100.00\%\\
      commons-lang & 62.50\% & 50.00\% & 12.50\% & 37.50\% & 25.00\% & 25.00\% & 62.50\% & 62.50\% & 62.50\% & \phantom{0}87.50\% & \phantom{0}87.50\% & \phantom{0}87.50\%\\
      dddlib & 37.50\% & 37.50\% & 12.50\% & 12.50\% & 12.50\% & 12.50\% & 37.50\% & 37.50\% & 37.50\% & \phantom{0}25.00\% & \phantom{0}25.00\% & \phantom{0}25.00\%\\
      datumbox & 22.22\% & \phantom{0}0.00\% & \phantom{0}0.00\% & 22.22\% & 22.22\% & 22.22\% & 44.44\% & 33.33\% & 33.33\% & \phantom{0}77.78\% & \phantom{0}77.78\% & \phantom{0}77.78\%\\
      openmrs-core & 37.93\% & 24.14\% & \phantom{0}6.90\% & 31.03\% & 24.14\% & 20.69\% & 44.83\% & 31.03\% & 24.14\% & \phantom{0}82.76\% & \phantom{0}65.52\% & \phantom{0}65.52\%\\
      basex & 47.83\% & 34.78\% & \phantom{0}8.70\% & 26.09\% & 26.09\% & 23.91\% & 52.17\% & 41.30\% & 36.96\% & \phantom{0}97.83\% & \phantom{0}84.78\% & \phantom{0}78.26\%\\
      \midrule
      Average & 43.69\% & 31.07\% & 7.77\% & 30.10\% & 24.27\% & 22.33\% & 49.52\% & 39.82\% & 35.92\% & \phantom{0}\textbf{85.44}\% & \phantom{0}\textbf{74.76}\% & \phantom{0}\textbf{71.84}\%\\
      \bottomrule
  \end{tabular}
  \label{tab:dynamic_performance}
\end{table}

\subsubsection{RQ4: Ablation Studies}

We conduct multiple ablation studies to evaluate the impact of the following parameters on \tool. 

\begin{itemize}
    \item Choice of LLMs: We evaluate the performance of \tool with gpt-3.5-turbo-0125, gpt-4-0125-preview, gpt-4-turbo-2024-04-09 and gpt-4o-2024-05-13 to understand the impact of different LLMs.
    \item Temperature: We evaluate the performance of \tool with different temperatures (ranging from 0 to 1), which controls the randomness of the generated response from the LLMs.
    \item Top P: We evaluate the performance of \tool with different values for nucleus sampling (ranging from 0 to 1) that is used to control the diversity of the generated text. Note that a Top P value of $x$ causes the LLMs to generate text using tokens with probabilities higher than $x$.
    \item  RAG and Dynamic Validation: We compared two variants without RAG, i.e., zero-shot, and without dynamic validation.
    \item Iteration number for the update phase: The dynamic validation process will keep executing until generating valid test code or reaching a pre-defined cutoff. By default, we limit the iteration number to 8. We aim to evaluate whether this threshold has a significant impact on \tool.
\end{itemize}

\begin{figure}[ht]
  \begingroup
  \begin{subfigure}[b]{0.3\columnwidth}
    \includegraphics[width=\textwidth]{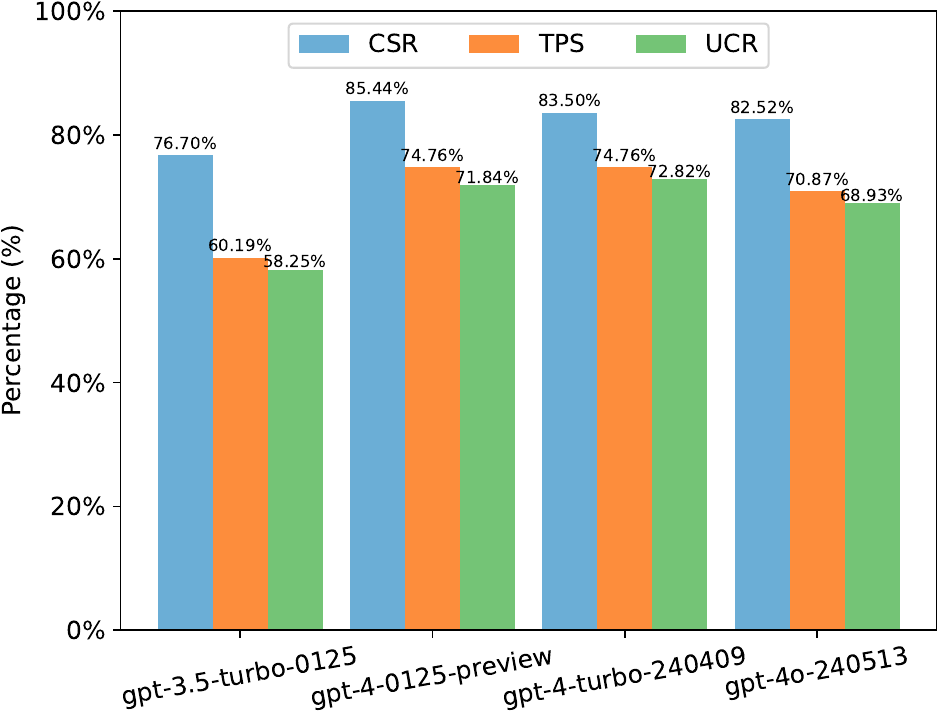}
    \caption{Effect of LLMs}
    \label{fig:llm}
  \end{subfigure}
  \begin{subfigure}[b]{0.3\columnwidth}
    \includegraphics[width=\textwidth]{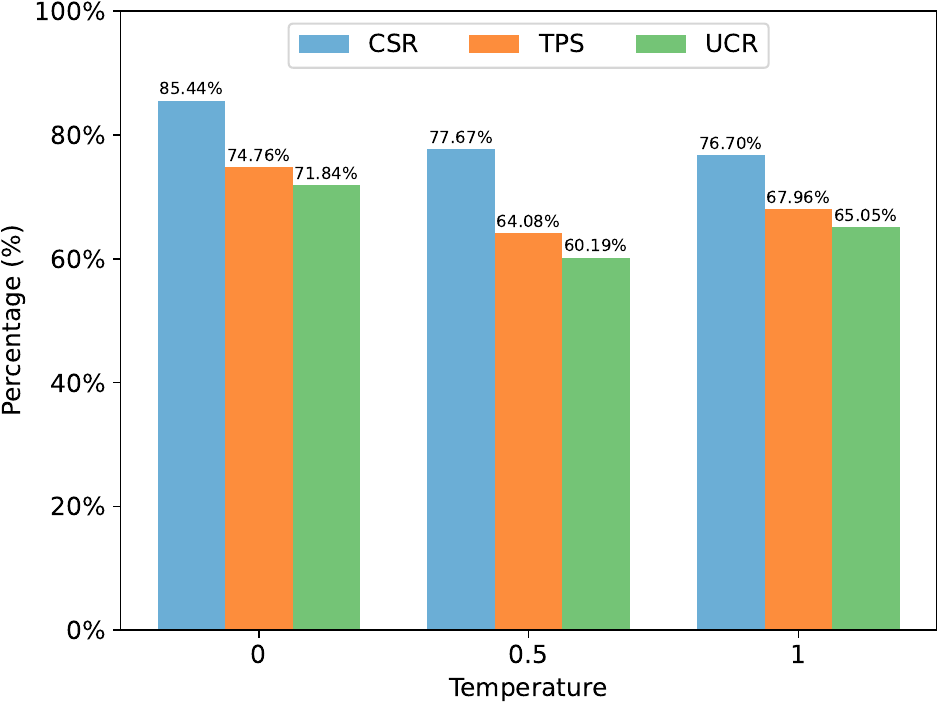}
    \caption{Effect of Temperature}
    \label{fig:temperature}
  \end{subfigure}
  \begin{subfigure}[b]{0.3\columnwidth}
    \includegraphics[width=\textwidth]{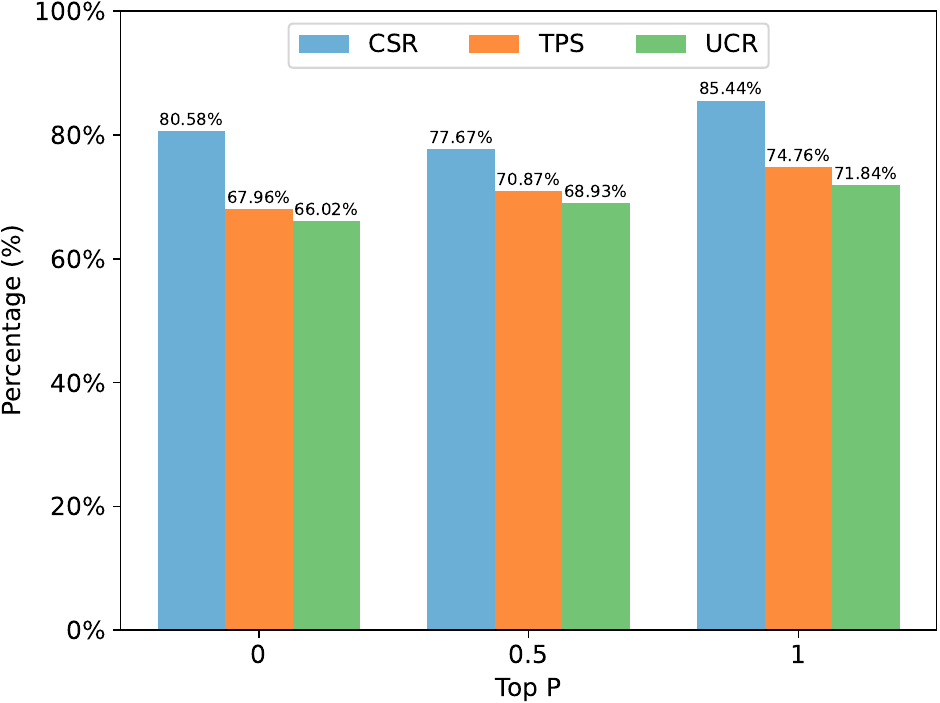}
    \caption{Effect of Top P}
    \label{fig:top_p}
  \end{subfigure}
  \medskip
  \begin{subfigure}[b]{0.3\columnwidth}
    \includegraphics[width=\textwidth]{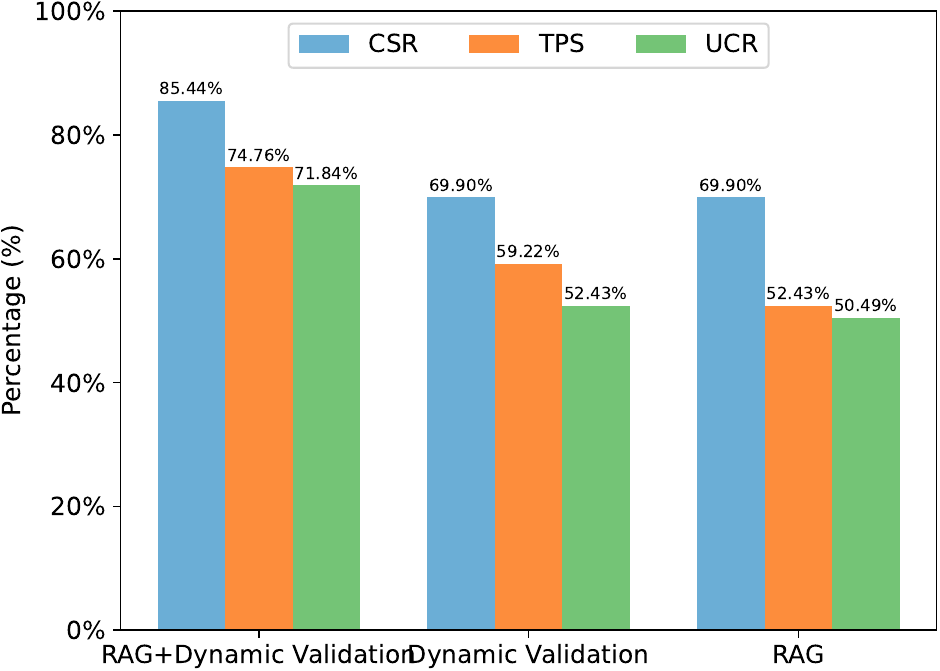}
    \caption{Effect of RAG and Dynamic Validation}
    \label{fig:rag_react}
  \end{subfigure}
  \begin{subfigure}[b]{0.6\columnwidth}
    \centering
    \includegraphics[width=0.7\textwidth]{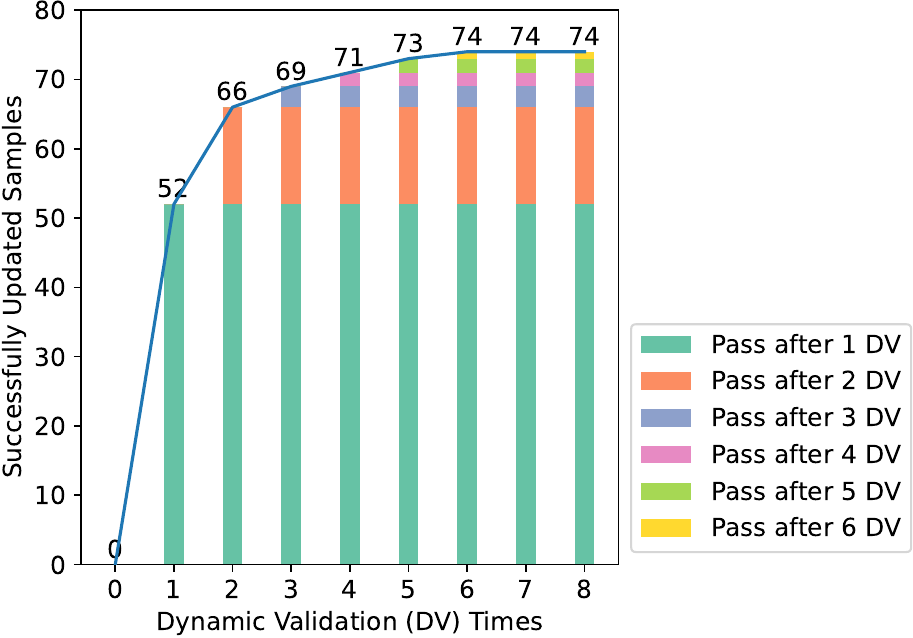}
    \caption{Times of dynamic validation}
    \label{fig:dynamic_validations}
  \end{subfigure}
  \endgroup
\caption{Ablation study results}
\label{fig:ab}
\end{figure}

For simplicity, we evaluate the impact of one variable at a time. \cref{fig:ab} summarizes the ablation study results respectively. Based on the results, several observations can be made.

First, GPT4 has a large improvement over GPT3.5, and the three versions of GPT4 have small internal differences. Specifically, gpt-4-0125-preview improves CSR, TPS and UCR by 8.74\%, 14.57\% and 13.59\% over gpt-3.5-turbo-0125.
The performance of gpt-4-0125-preview and gpt-4-turbo-2024-04-09 is almost equal.
Surprisingly, the latest model, gpt-4o-2024-05-13, showed a decrease of 2.92\%, 3.89\%, and 2.91\% in CSR, TPS, and UCR respectively, compared to gpt-4-0125-preview.

Secondly, the CSR and TPS of the updated test code decrease as the temperature increases, with a peak at a temperature of 0.
UCR is lowest when the temperature is 0.5.

Thirdly, a higher value of Top P improves the performance of \tool.
When the Top P value is low, the LLM maintains fewer candidates for the output, resulting in more consistent but lower-quality updates to the test code.

Fourthly, utilizing RAG by providing an example to the LLM enhances the relevance of its output to code changes, significantly improving CSR, TPS, and UCR.
Furthermore, incorporating dynamic validation results to guide the LLM in optimizing low-quality test code also contributes to substantial improvements in these metrics.

Lastly, in order to understand the impact of the number of iterations, we analyze the 74 co-evolution samples that are successfully updated by \tool.
The statistics are shown in~\cref{fig:dynamic_validations}. 
We observed that in 70.27\% of cases, the test code was successfully updated in the first iteration, indicating that most updates do not require feedback from the test environment to be completed.
Among the remaining 29 cases, 22 were successfully updated in subsequent iterations.
Within these 22 cases, 63.63\% required dynamic validation twice, meaning the update was completed after collecting test information only once.
This demonstrates that \tool can effectively update the test code by collecting feedback from the test environment.
Furthermore, the dynamic validation process reduces the LLM's ‘hallucination’ phenomenon by 75.86\% and improves update effectiveness by 42.31\%.

%% file: chapters/discussion.tex
\subsection{Discussion}

It was observed that \tool's performance on \emph{dddlib} is not as strong as expected. 
After an in-depth investigation on \emph{dddlib}, it turns out that the root cause of such performance downgrade lies in the unsuccessful updated samples.
The \emph{dddlib} project contains two successfully configured commits with four samples each.
In commit \emph{32f7a3d1d6d33c31e154fe893ab8ea3b73b32389}, out of the four samples, only the first sample is successfully updated, while the remaining three contain compilation errors. Upon analyzing these samples, we found that all samples have similar modified structures but different identifiers, i.e., variable names and method names used in the test cases, leading to the retrieval of a suboptimal sample.
A similar result occurs when \tool tackling the other commit \emph{566855c10bb0d875e493ef0c5d4d075c80c20a78}, in which one out of the four samples is correctly updated.
Therefore, in the RAG process for the update task, leveraging similarities in code and modification structures may be more effective than relying solely on textual similarity.

\begin{figure}[t]
	\centering	 
        \includegraphics[width=1\columnwidth]{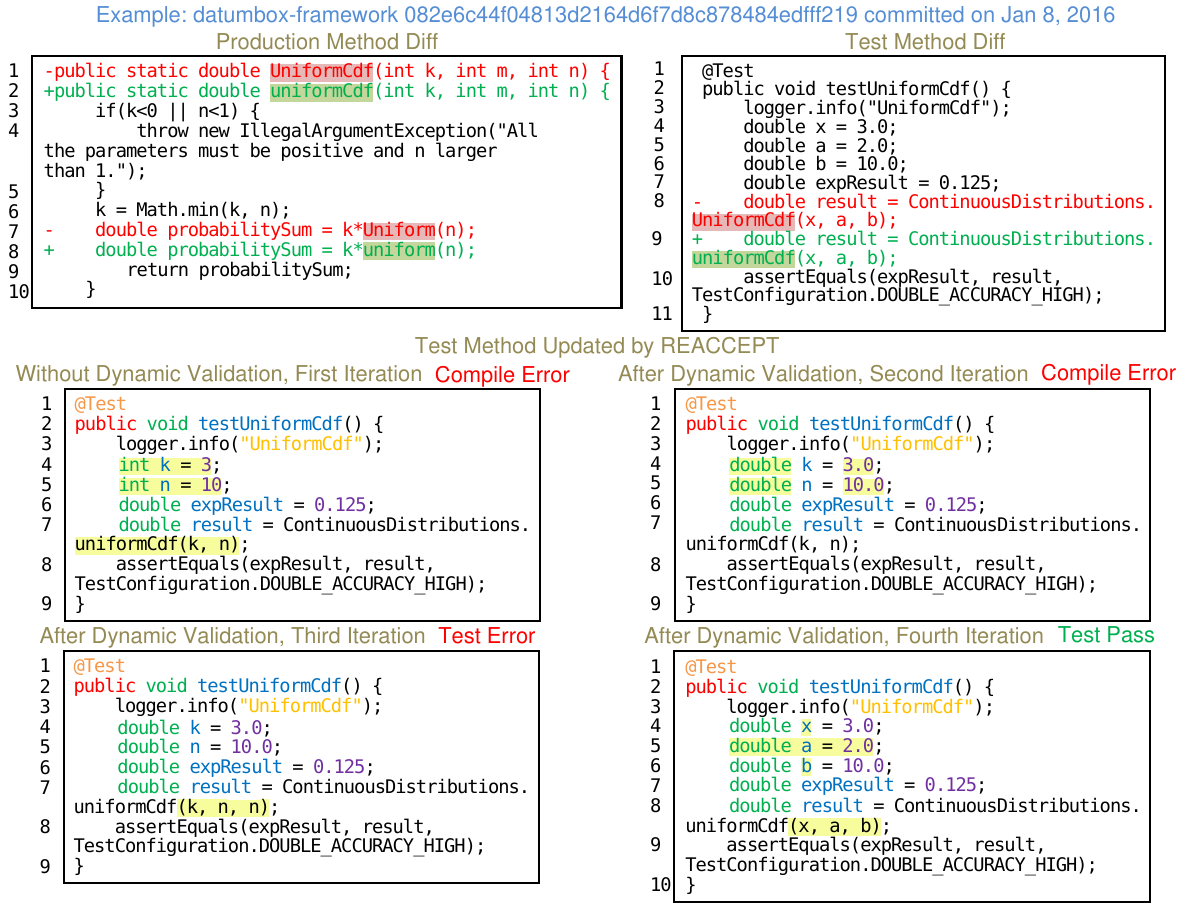}
	\caption{Impact of dynamic validation on updating obsolete test code}
	\label{fig:dynamic_validation2}
\end{figure}

To better clarify the effect of dynamic validation in optimizing the test code generation, we conducted a case study on a successfully updated sample from \emph{datumbox-framework}.
As shown in \cref{fig:dynamic_validation2}, developers renamed the method in the production code, resulting in a compilation error when launching the existing test code.
To fix such an error, the test code must be refactored to invoke the method through the correct method name \textit{uniformCdf}.
After \tool identified this obsolete test code, it proceeded to the updating task using the LLM.


In the first iteration, the LLM renamed the \textit{UniformCdf} method called in the test code.
However, due to the influence of the non-changed part of the production code, the LLM incorrectly called the \textit{uniformCdf} method, reasoning a Java compilation error and inconsistent semantics compared to the original one.

In the second iteration, the LLM accepted the results of the dynamic validation and modified the actual argument type of \textit{uniformCdf}.
But it didn't modify the length of the actual argument list, the compilation error still occurred.

In the third iteration, the LLM attempted to pass an extra \textit{n} into the actual argument list and the compilation was successful.
However, when executing the test, a test error was raised for the formal argument \textit{a} in the \textit{uniformCdf} method needed to be less than \textit{b}.

In the fourth iteration, the LLM finally modified the formal argument when calling \textit{uniformCdf} correctly so that the test code can be executed successfully.

This example highlights the advantages of using dynamic validation with an LLM to update test code, significantly enhancing both the quality and practical performance of the test code.

\subsection{Threats to Validity}

\subsubsection{Internal Validity}
Although we've summarized general experience to guide the LLM, its quantity and quality are still not comprehensive enough. In complex or unique cases, this experience may not cover all possible code change scenarios. Expanding and enhancing the quality of this experience, along with providing more detailed classifications for different types of code changes, could further improve identification accuracy.

In addition, despite utilizing knowledge retrieval and dynamic validation to guide LLM in updating and optimizing the obsolete test code, there are still some update failures.
We use code diff as the code change representation, which is LLM-friendly. However, during knowledge retrieval, focusing solely on textual similarity may overlook code structure and change types. Introducing a vector representation of AST differences for a more fine-grained, structured search could enhance update effectiveness.

\subsubsection{Data Validity}
We collected co-evolution samples from SITAR, CHOSEN, and CEPROT, mined using heuristic rules. 
Upon reviewing some code change pairs, we found that the production and test code only have dependencies, with uncorrelated changes and mismatched class or method names. These production code changes are insufficient for the LLM to effectively update the obsolete test code, resulting in low test coverage.

\subsubsection{External Validity}
The LLM used in our experiments is gpt-4-0125-preview. While it has moderate NLP and code understanding capabilities, its performance is average in the current competitive LLM landscape. 
We believe that using more advanced models could significantly improve identification accuracy and effectiveness.


%% file: chapters/relatedworks.tex
\section{Related Work}\label{sec:related_work}
Previous work has proposed various techniques to model correlations between production code and test code, mainly focusing on mining co-evolution patterns~\cite{zaidman2008mining, van2009establishing, lubsen2009using, pinto2012understanding, marsavina2014studying, sohn2022using, shimmi2022leveraging, kitai2022have, sun2023revisiting, cai2024fly}. 
Commonly, these methods define heuristic rules to create traceability links. 
Zaidman et al.~\cite{zaidman2008mining} were among the first to study PT co-evolution patterns in two open-source projects, using software visualization techniques to illustrate co-evolution trends.
White et al. \cite{white2020establishing} proposed TCtracer, which automatically establishes traceability links between production and test code at the method and class levels.
Huang et al. \cite{huang2024towards} introduced a new approach that gauges the likelihood of co-evolution through extracted code changes, code complexity, and certain semantic features. 


In recent years, numerous studies on identifying and updating PT co-evolution relations have emerged~\cite{palomba2016automatic, wang2024testeval, piya2023llm4tdd, shin2024domainadaptationcodemodelbased}.
Wang et al. \cite{wang2021understanding} studied the factors influencing test code updates. The proposed SITAR model uses historical PT co-evolution data to identify obsolete test code.
Liu \cite{liu2023drift} proposed a machine-learning-based method called Drift, which predicts obsolete test cases at the method level.
Hu \cite{hu2023identify} introduced a novel deep-learning-based method, CEPROT, to identify obsolete test cases and automatically update them. 
Sun et al. \cite{sun2023revisiting} explored the validity of common assumptions for collecting and labeling PT co-evolution samples. 
They also proposed a two-stage strategy-based method, CHOSEN, for identifying PT co-evolution.
Yaraghi et al.~\cite{yaraghi2024automated} proposed a method called TARGET that utilizes pre-trained models in conjunction with program testing to repair broken test cases.
TARGET may be applicable to PT co-evolution problems, but we did not find an open-source implementation of TARGET to compare with \tool{}.
In addition, TARGET only partially validates the generated test cases by compiling and executing them, while our approach considers more metrics related to the quality of test code.

%% file: chapters/conclusion.tex
\section{Conclusion}\label{sec:conclusion}

In this work, we propose \tool{}, a novel approach that leverages LLMs and dynamic validation to fully automate PT co-evolution (i.e., capable of identifying and updating obsolete test cases). 
The evaluation results on a dataset of 537 collected Java projects show that \tool achieves significant improvements. 
In the future, we intend to investigate additional strategies to improve \tool, such as integrating contextual information or improving code change representation learning.